# Upconversion nonlinear structured illumination microscopy

*Baolei Liu[1,2], Chaohao Chen[1], Xiangjun Di[1], Jiayan Liao[1], Shihui Wen[1], Qian Peter Su[1], Xuchen Shan[1], Zai-Quan Xu[2], Lining Arnold Ju[3,4], Fan Wang[1]\*, Dayong Jin[1,5]\**

[1] Institute for Biomedical Materials & Devices (IBMD), Faculty of Science, University of Technology Sydney, Sydney, NSW 2007, Australia

[2] School of Mathematical and Physical Sciences, Faculty of Science, University of Technology Sydney, Ultimo 2007, Australia

[3] School of Biomedical Engineering, Faculty of Engineering and Charles Perkins Centre, The University of Sydney, Camperdown, NSW, Australia 2006

[4] Heart Research Institute, Newtown, NSW, Australia 2042.

[5] UTS-SUStech Joint Research Centre for Biomedical Materials & Devices, Department of Biomedical Engineering, Southern University of Science and Technology, Shenzhen, China

**Abstract:** Video-rate super-resolution imaging through biological tissue can visualize and track biomolecule interplays and transportations inside cellular organisms. Structured illumination microscopy allows for wide-field super resolution observation of biological samples but is limited by the strong absorption and scattering of light by biological tissues, which degrades its imaging resolution. Here we report a photon upconversion scheme using lanthanide-doped nanoparticles for wide-field super-resolution imaging through the biological transparent window, featured by near-infrared and low-irradiance nonlinear structured illumination. We demonstrate that the 976 nm excitation and 800 nm up-converted emission can mitigate the aberration. We found that the nonlinear response of upconversion emissions from single nanoparticles can effectively generate the required high spatial frequency components in Fourier domain. These strategies lead to a new modality in microscopy with a resolution of 130 nm, 1/7th of the excitation wavelength, and a frame rate of 1 fps.

INTRODUCTION

Fluorescence microscopy has been widely used to visualise cellular structures, biomolecular distributions and biological processes[1,2]. However, many subcellular structures, organelles and molecular analytes are typically smaller than a few hundreds of nanometres, which cannot be resolved by conventional microscopy due to the optical diffraction limit. Super-resolution microscopy techniques, including stimulated emission depletion (STED) microscopy[3], single molecule localization microscopy[4,5], super-resolution optical fluctuation imaging (SOFI)[6] and structured illumination microscopy (SIM)[7,8,9], have been developed to bypass this limitation.

SIM typically requires closely spaced periodic patterns to down-modulate the high spatial frequency information in the sample so that with the support of optical transfer functions the high frequency information can be reconstructed from a series of images obtained from the patterned illuminations at various orientations. SIM typically offers the high-speed imaging with a resolution at around ¼ of the excitation wavelength. When the high excitation power is used, the nonlinear saturated photo-response can be explored to further improve the resolution of SIM in the regime of 50 nm[10] and resolve sub-cellular structures[11]. New advances made in de-noising process and modified excitation conditions have been applied to SIM, so that nonlinear SIM[10,11,12], Hessian-SIM[13] and grazing incidence SIM[14], have been recently developed with high imaging speed, for the observations of ultrastructures of cellular organelles and their structural dynamics, such as mitochondrial cristae[13].

The next challenge is to explore the possibilities in using these techniques for thick tissue neuroscience imaging and nanomedicine tracking, as the strong scattering and absorption aberrate the structured illumination patterns and introduce unwanted out-of-focus light, both deteriorating the imaging resolution[15]. To address this challenge, near-infrared excitation has been implemented to mitigate the aberration. Two-photon[16] or multi-photon excitation[17] in conjugation to spot scanning SIM have been reported to improve the imaging depth through tissue, but at the price of low speed caused by the spot scanning scheme. Organic fluorescent dyes and proteins are the most common imaging probes for SIM, because of their outstanding staining and specific targeting ability to organelles[8,13,18,19]. Nevertheless, these probes require the tightly focused and high-power pulsed laser to activate the multi-photon absorption process, due to their small multiphoton absorption cross-section[20]. The required high excitation power, especially by nonlinear SIM, limits long-duration visualization of sub-cellular structure in living cells.

Upconversion nanoparticles (UCNPs) is an emerging near-infrared (NIR) imaging probe for optical imaging. UCNPs typically consist of multiple lanthanide sensitizer ions to absorb NIR photons and activator ions to emit light, both can be highly doped in fluoride nanocrystal host. Hence, these nanoparticles can effectively absorb NIR photons and convert them into ultraviolet, visible and NIR photons, through nonlinear energy transferring process[21,22]. Benefiting from its unique nonlinear response, non-bleaching, non-blinking photo-stability and anti-Stoke excitation-emission properties[23,24,25,26,27], UCNPs have been recently discovered as a new library of super resolution imaging[28,29] and single molecule tracking probes [23,27,30]. We have recently demonstrated a near-infrared emission saturation nanoscopy method that can detect single nanocrystals through 93 μm thick tissue with a 50 nm resolution[29,31]. However, these super resolution imaging modalities require focused donut beam excitations and suffer from the low scanning rate, therefore not suitable for video rate super resolution imaging.

Here we report a strategy of upconversion nonlinear SIM (U-NSIM) towards video-rate super-resolution imaging through biological tissues. We apply Ytterbium ($Yb^{3+}$) and Thulium ($Tm^{3+}$) co-doped UCNPs as the imaging probe that emits up converted NIR emissions at 800 nm upon the NIR excitation at 980 nm, both within the transparent biological window, so that to extend the imaging penetration depth. Setting the 980 nm excitation structured pattern can not only mitigate the aberration on pattern but also easily activate UCNPs to emit bright NIR photons. The unique nonlinear photo-response of UCNPs enables the onset of an efficient nonlinear mode on SIM for obtaining higher-frequency imaging resolution. We further demonstrate that fine tuning of the doping concentration in UCNPs can modify the nonlinearity of the photon-response towards the further improved optical resolution of $1/7^{th}$ of the excitation wavelength.

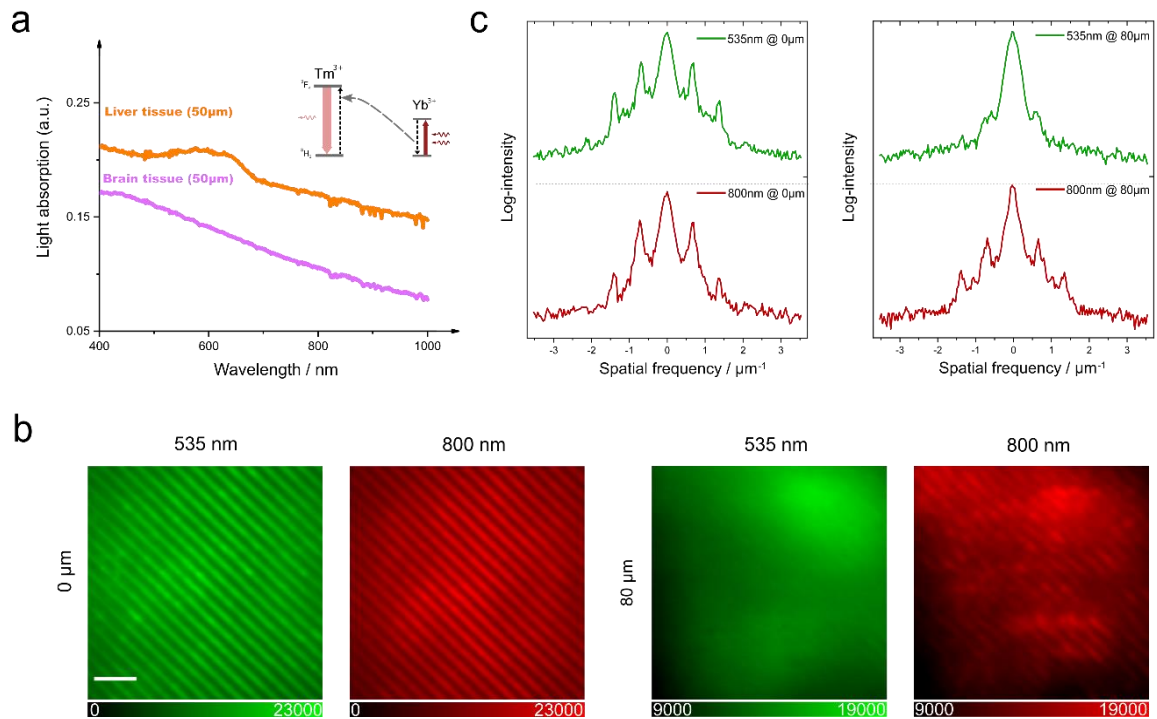

**Figure 1** The scheme and advances of upconversion nonlinear SIM for super resolution imaging through thick tissues. **a** Light absorption spectra of 50 μm thick mice brain and liver tissues, measured by a commercial UV-Vis spectrophotometer. Single lanthanide-doped upconversion nanoparticle (UCNP) with a network of thousands of co-doped ytterbium ions ($Yb^{3+}$) as sensitizers and thulium ($Tm^{3+}$) ions for activators can absorb 976nm excitation photons and convert them into 800 nm emissions with nonlinear photo-response. **b** Under the sinusoidal structured excitation at 976 nm, the comparison fluorescence images collected through the emission bands of 535 ± 25 nm and 808 ± 10 nm with and without 80 μm-thick brain tissues. Scale bar: 5 μm. **c**. Line profiles of Fourier spectra (on a logarithmic scale) of the cross-section profiles in **b**. For comparison purpose, the emission intensities from the same sample area through the bands of 535nm and 800 nm have been adjusted to reach the same level by tunning the excitation power.

RESULTS AND DISCUSSION

To evaluate the upconversion strategy to achieve video-rate super-resolution imaging through thick tissues, we first examine the light penetration depths of UCNPs' multi-colour emissions. Under the low continuous wave (CW) excitation condition (around 10 kW/cm$^2$), UCNPs can effectively convert the energy from 976nm photons into the two-photon state $^3H_4$ that emits 800 nm photons (Supplementary Note 3 and Figure S2), where the tissue has less absorption of its emission. Figure 1a shows the light absorption spectra of a mice brain tissue slice and a liver tissue slice with 50 μm thickness, indicating 800 nm emission and 976 nm excitation have better penetration ability than that at visible wavelength range. Figure 1b shows the comparison penetration ability of structured patterns at 535nm and 800nm emission bands. In this experiment, we mix two types of UCNPs (NaYF4: 2% $Er^{3+}$, 20% $Yb^{3+}$ and NaYF4: 2% $Tm^{3+}$, 20% $Yb^{3+}$) and uniformly disperse them onto a glass slide (Supplementary Note 1 & 2). Then we use a 976 nm sinusoidal structured pattern to excite the layer of UCNPs and image the emission patterns (Supplementary Note 4 and Figure S4). Without the tissue, marked as *0 μm*, both 535 nm (from $Er^3$) and 800 nm (from $Tm^{3+}$) emissions produce the desirable structured emission patterns. When an 80 μm tissue slice is placed on the top of the UCNPs layer, the 535 nm emission pattern is heavily distorted and almost lost its structure information, while the 800 nm emission pattern mitigates the scattering and well keeps its pattern. We further quantify the information that preserved in the emission patterns by Fourier domain image analysis (Figure 1c). Both the diagonal cross sections of Fourier spectra of the emission patterns from two bands at 0 *μm* show clear peaks, indicating a sufficient ability to transfer designed spatial frequencies. However, the 535 nm emission pattern loses this spatial information after an 80 μm tissue, while the 800 nm emission pattern well keeps the spatial information. This indicates that the upconversion SIM method using NIR emission bands is superior to visible emissions as it can send and detect structured excitation and emission patterns through the thick tissue.

SIM operated in the linear response regime of the fluorescence probes generally enhances the resolution by a factor of approximately 2, compared with that in the wide-field microscopy. Nonlinear operation of SIM requires high excitation power density for the probes to produce nonlinear photo-responses, e.g. fluorescence saturation[10] and fluorophores depletion[11,12], to surpass the resolution limitation of linear SIM. Nonlinear SIM can theoretically produce an unlimited factor of enhancement on the imaging resolution[10]. The photon upconversion process is a typical nonlinear process as it requires multi-photon excitation to emit the multi-colour upconversion emission from a ladder-like multi-metastable excited states of rare earth ions. We first measure the 800 nm emission saturation curves from single nanoparticles (Supplementary Note 1, 2 & 3 and Figure S3). Figure 2a shows the obvious nonlinear photo-response behaviours of UCNPs, and the nonlinearity (the rising-up slope) increases with emitters' doping concentration. Higher concentration increases the energy transferring rate from $Yb^{3+}$ to $Tm^{3+}$ and boosts the cross-relaxation between $Tm^{3+}$ ions, thereby enhancing the nonlinearity on photo-response. However higher doping concentration often leads to lower emission intensity if lower excitation power, e.g. 4 kW·cm$^{-2}$ is used in SIM (Figure 2b). We therefore select the doping concentration of 4% to compromise between high nonlinearity and high brightness for an optimised imaging quality. We find that the emission intensity can be further improved by tunning the doping concentration of $Yb^{3+}$ ions to 40% (Figure 2c), as an optimized $Yb^{3+}$ concentration can increase energy transferring rate from $Yb^{3+}$ to $Tm^{3+}$. Figure 2d reveals the resolution-resolving power of nonlinearity for UCNPs with different concentration by measuring the amplitude of high-frequency harmonics ($H$) in the Fourier transform of their emission pattern (Figure 2d, insert) under a sinusoidal excitation pattern. Compared with diffraction-limited wide-field imaging result under uniform illumination ($H = 0$) and linear SIM ($H = 1$), the additional harmonics in upconversion NSIM ($H \geq 2$) allow for a lateral resolution to $\sim\lambda/[2NA(H+1)]$, where *NA* stand for the numerical aperture[10-12]. Clearly, UCNPs with higher $Tm^{3+}$ concentration have stronger harmonic peaks, with $H = 3$ for 2% & 4% $Tm^{3+}$ and $H = 4$ for 8% $Tm^{3+}$. Here we identify the 4% $Tm^{3+}$ 40% $Yb^{3+}$ co-doped UCNPs are the best suitable probes for upconversion NSIM, as it provides $H = 2$ harmonics whilst under the lower excitation power than that with 8% $Tm^{3+}$ to offer improved control of photo toxicity.

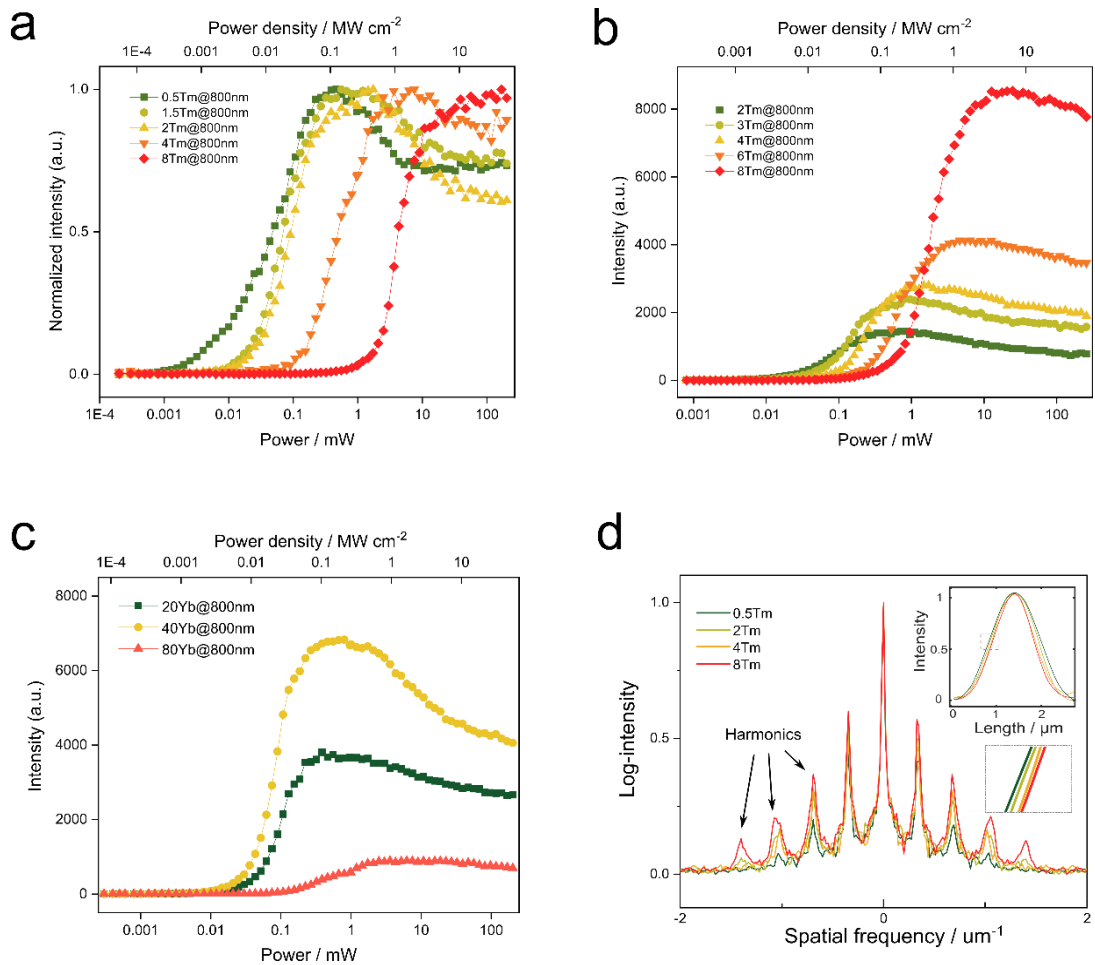

**Figure 2** The saturation intensity curve of the 800 nm emissions from single UCNPs. **a**, Normalized emission saturation curve to the maximum intensity of a single UCNP (NaYF4: 20% $Yb^{3+}$, x% $Tm^{3+}$ nanoparticles, x = 0.5, 1.5, 2, 4 and 8) under 980 nm excitation. **b**, 800 nm emission saturation curves obtained for a single UCNP (NaYF4: 20% $Yb^{3+}$, x% $Tm^{3+}$ nanoparticles, x = 2, 3, 4, 6 and 8). **c**, 800 nm emission saturation curves obtained for a single UCNP. (NaYF4: x% $Yb^{3+}$, 4% $Tm^{3+}$ nanoparticles, x = 20, 40 and 80) **d**, Fourier transforms

corresponding to the intensity profiles measured with different doped UCNPs, and the insert (right middle) is the cross-section profiles of measured emission patterns (upper right insert).

We further examine the resolving power of upconversion NSIM by resolving single UCNPs on glass slide (Figure 3). With the schematics of the optical system shown on Supplementary Figure S4, a DMD is used to generate the excitation pattern and a 60x water immersion objective (NA = 1.27) is used to direct excitation and collect the emission (Supplementary Note 4). Figure 3a shows a typical upconversion-NSIM image compared with the diffraction-limited wide-field excitation image across the field of view (FOV) of 32.3- by 32.3-μm, with the comparison images of the 8- by 8-μm area (orange square) by wide-field (WF) microscopy (Figure 3b), Wiener deconvoluted result of WF (Figure 3c), upconversion linear SIM (U-LSIM, Figure 3d) and upconversion - NSIM (U-NSIM, Figure 3e). To quantify the resolving power by different modalities, Figure 3f shows the comparison result in resolving a pair of UCNPs separated with 320 nm, by WF, deconvolution, linear SIM and upconversion NSIM (Supplementary Figure S5). Here, U-NSIM (1) is reconstructed from 15 raw frames (3 orientations and 5 phase shifts), and U-NSIM (2) requires 25 raw frames (5 orientations and 5 phase shifts). U-NSIM (2+) stands for NSIM with Richardson–Lucy deconvolution (iteration =10) process. By imaging another 2.9- by 6.5-μm area (green rectangle) Figure 3h shows that the NSIM modality can resolve two adjacent nanoparticles with spacing of 161nm with a typical full width at half maximum (FWHM) about 130 nm ($\sim \lambda_{exc}/7.5$), measured by Gaussian fitting of the profiles.

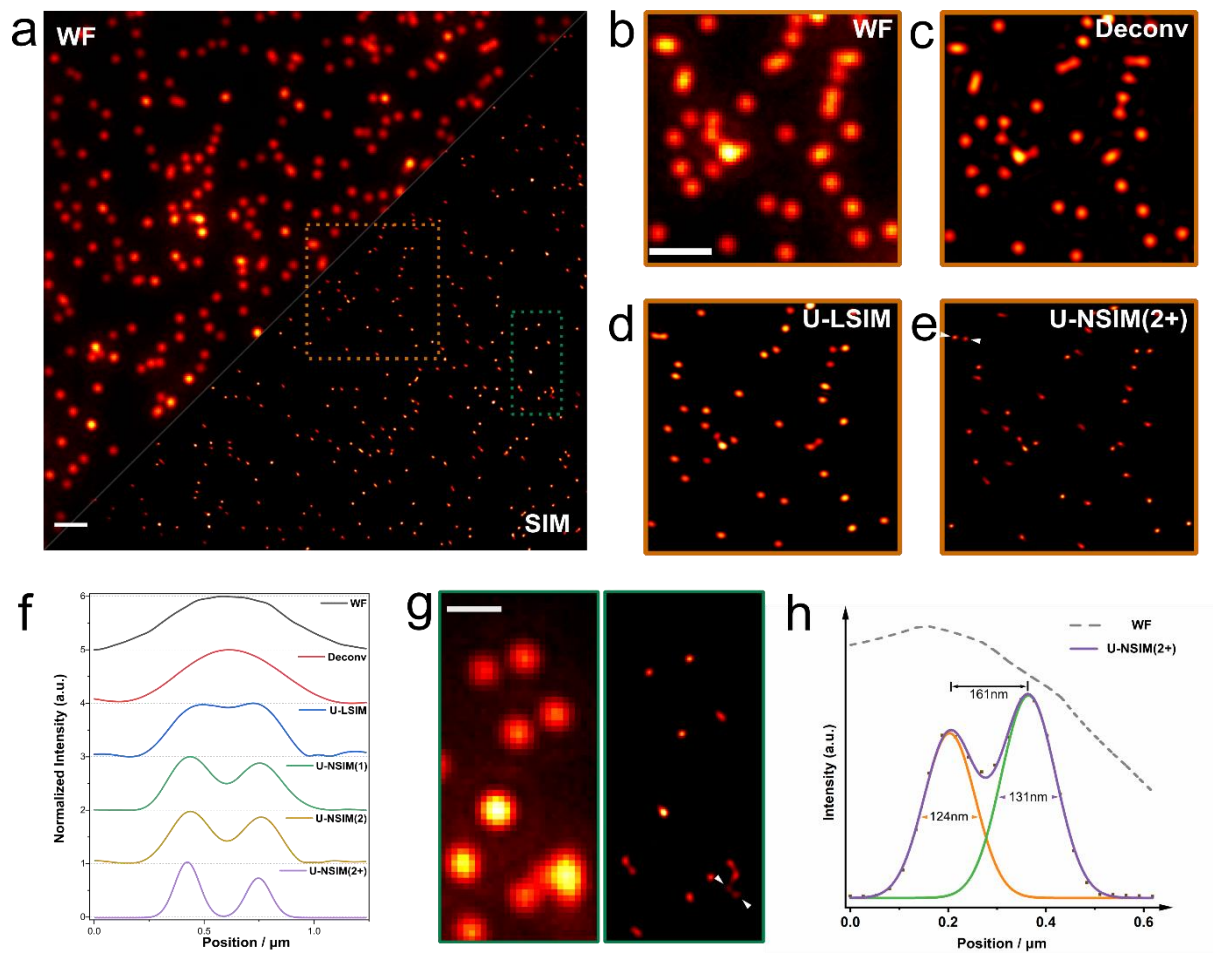

**Figure 3** Super-resolution imaging reconstructions of single upconversion nanocrystals. **a**, Wide-field (left) and super-resolution (right) images of the 4% Tm-doped UCNPs. **b-e**, Comparison imaging results of a selected area (orange frame) with different methods, **b**: wide-field microscopy; **c**: Wiener deconvolution; **d**: linear SIM; **e**: nonlinear SIM. Scale bar: 2 μm. **f**, Line profiles of the resolved particles in upper-left corner of **e** with a range of imaging methods. **g**, Comparison imaging results of the green framed area. **h**, Line profiles of lower right corner in **g**. Scale bar: 1 μm. The diameter of UCNPs is 40 nm (confirmed by TEM). The reconstruction processes are performed by both fairSIM[32] in imageJ[33] and a written MATLAB code.

CONCLUSION

In conclusion. We demonstrate that the upconversion-NSIM strategy as a new modality for in-depth super resolution imaging. We find that the nonlinear photo-response properties of UCNPs can produce a series of additional harmonics with higher spatial frequencies, suggesting a new scope for probe developments. The obvious advances breakthrough by NSIM using UCNPs is the improved imaging resolution at low excitation power of CW laser for sub-cellular dynamic tracking of single nanoparticles through deep tissue. Compared with our recent works on single point scanning nanoscopy[29,31], the U-NSIM offers high frame rate towards video rate imaging. Notably, this strategy can be directly adapted to enhance the imaging penetration depth in light sheet based SIM[34], adaptive optics[35,17], and reconstruction algorithms[36]. Though specific labelling of UCNPs to the subcellular structures in live cells remain as a challenge, the recent progress on functionalization of UCNPs show the promise in using the labelling and single particle tracking to release multimodality and multiplexing super-resolution imaging of biomolecules of interests in live cell environment[21,30].

## Supporting Information

The supporting information includes the Material and Methods section and supporting figures 1 through 5.

## Corresponding Author

* fan.wang@uts.edu.au; dayong.jin@uts.edu.au

## Author contributions

F.W., B.L. and D.J. conceived the project and designed experiments. B.L., X.S. and C.C. conducted the optical setup and performed the optical experiments. J.L. and S.W. synthesized the upconversion nanoparticles. X.J. performed the nanoparticles modification. B.L. and Z.X. performed sample preparation. Q.S. and L.A.J. helped prepare the biological samples. B.L.,

F.W. and C.C. analysed the results. B.L., F.W and D.J. wrote the paper with input from the other authors.

**Acknowledgements**

This work was supported by grants from the Australian Research Council Discovery Project (DP190101058 - F.W, DP200101970 - L.A.J, Q.P.S.) and Sydney Research Accelerator prize (SOAR – L.A.J). F.W. is an Australian Research Council DECRA fellow (DE200100074). L.A.J. is an Australian Research Council DECRA fellow (DE190100609). The authors have no conflicting interests to declare.